\documentclass[9pt,conference]{IEEEtran}

\usepackage{bm}
\usepackage{dcase2026,amsmath,graphicx,url,times,booktabs,tabularx}
\graphicspath{{../}{./}}

\makeatletter
\renewcommand\addcontentsline[3]{}
\makeatother

\usepackage[T1]{fontenc}
\usepackage{mathptmx}   % Times for both text and math (numbers in Times)
\providecommand{\team}[1]{{\fontfamily{lmtt}\selectfont #1}}
\providecommand{\sys}[1]{{\fontfamily{lmtt}\selectfont #1}}
\providecommand{\pct}[1]{#1\,\%}
\providecommand{\pp}[1]{#1\,pp}
\providecommand{\bn}[1]{#1\,B}

\renewcommand{\footnoterule}{%
  \kern -3pt
  \hrule width 0.4\columnwidth height 0.4pt
  \kern 2.6pt}

\title{Summary of DCASE 2026 Task 5: Audio-Dependent Question Answering}

\name{\begin{tabular}[t]{@{}c@{}}
Haolin He,$^{1,\ast}$
Renhe Sun,$^{2,\ast}$
Zheqi Dai,$^{1}$
Xingjian Du,$^{3}$
Chunyat Wu,$^{1}$
Zining Liang,$^{1}$\\
Zhengxi Liu,$^{1}$
Jiahe Lei,$^{1}$
Runbang Wang,$^{1}$
Jiayi Zhou,$^{2}$
Mingru Yang,$^{2}$
Xiquan Li,$^{4}$
Yun Chen,$^{5}$\\
Xie Chen,$^{4}$
Zhiyao Duan,$^{3}$
Weiqiang Wang,$^{2}$
Mark D. Plumbley,$^{5}$
Jian Liu,$^{2,\dagger}$
Qiuqiang Kong$^{1,\dagger}$%
\thanks{$^{\ast}$Equal contribution. $^{\dagger}$Corresponding authors.}%
\thanks{The authors thank the 14 participating teams of DCASE 2026 Task~5 for their submissions and technical reports.}
\end{tabular}}
\address{$^{1}$The Chinese University of Hong Kong, Hong Kong SAR \;
$^{2}$Ant Group, China \\
$^{3}$University of Rochester, USA \;
$^{4}$Shanghai Jiao Tong University, China \;
$^{5}$King's College London, UK}

\begin{document}

\maketitle

\begin{abstract}
DCASE~2026 Task~5 introduces Audio-Dependent Question Answering (ADQA), which tests whether large audio-language models answer from the audio rather than from textual priors. An Audio-Dependency Filtering (ADF) pipeline combines silent-audio probing, per-option perplexity, a large language model (LLM) commonsense check, and human review to remove items solvable from text alone. The 3000 items that pass form the ADQA-Bench evaluation set, spanning music, speech, and environmental audio. The inaugural edition draws 14 teams and 36 submissions across two tracks defined by total parameter count (up to 100B and under 10B). A Chung-Ang University ensemble of MOSS-Audio-8B-Thinking and Qwen3-Omni-30B reaches the top overall accuracy at \pct{58.33}, and a MOSS-only configuration from the same team leads the sub-10B track at \pct{57.30}. Across the 30 submissions with a comparable development score, evaluation accuracy falls by 11.91 percentage points (pp) on average (median 10.91\,pp) on the hidden evaluation split, which is designed to be harder than the development split. The most common building blocks are: the MOSS-Audio-8B-Thinking backbone (13 of 36 submissions), Low-Rank Adaptation (LoRA) fine-tuning on AudioMCQ-StrongAC, and preference or reinforcement-learning objectives---Group Relative Policy Optimization (GRPO) in five teams, Group reward-Decoupled Normalization Policy Optimization (GDPO) in two. At test time, prompt engineering is near-universal, and majority or choice-permutation voting is common. Every system misses the same set of 233 evaluation items.
\end{abstract}


\section{Introduction}
\label{sec:intro}

Large Audio-Language Models (LALMs) can inflate their accuracy on
multiple-choice audio benchmarks by partly exploiting textual shortcuts rather than the audio itself. Replace the audio with silence and
state-of-the-art LALMs still answer many questions correctly, using
linguistic priors and dataset
artifacts~\cite{he2025audiomcq,sakshi2025mmau,ma2025mmar,wang2025mmsu}.
Progress requires benchmarks that remove such
items, and models that answer from the audio signal itself.

DCASE~2026 Task~5, Audio-Dependent Question Answering
(ADQA)~\cite{he2025audiomcq}, addresses this gap. An
Audio-Dependency Filtering (ADF) pipeline drops questions that a
text-only model can solve, leaving a 3000-item evaluation set
(ADQA-Bench). Submissions are open-source LALMs that answer
four-way multiple-choice questions grounded in the audio. Two
tracks run in parallel, split by total parameter count: an Overall
track (${\le}\bn{100}$) and a~\mbox{Lightweight track ($<\bn{10}$)}.

The top Overall system, \team{Lim\_CAU} from Chung-Ang
University~\cite{Lim_CAU_t5_2026}, reaches \pct{58.33}; the same
team wins the Lightweight track at \pct{57.30}. The strongest
organizer baseline, Qwen3-Omni-30B-A3B-Instruct (hereafter
Qwen3-Omni-30B)~\cite{qwen3omni_2025}, scores \pct{62.48} on
development. Baselines are run only on development. Team
submissions average ${\sim}$\pp{12} below their own development
scores on the harder hidden split~\cite{he2025audiomcq}
(\cref{sec:dev_eval}), so this baseline number is not directly
comparable to the team eval accuracies. Every system beats the
random-guess baseline (\pct{25.46}, above \pct{25} likely due to
option-length and distractor artifacts). The median across the
36 submissions is \pct{50.87}, and the top
score of \pct{58.33} leaves clear~headroom~on~\mbox{ADQA-Bench}.

\Cref{sec:task} details the task and ADF, \cref{sec:overview} the
submissions, \cref{sec:methods} the methods, \cref{sec:analysis}
the performance analysis, and \cref{sec:discussion} the discussion and future directions.

\section{Task Setup and Dataset}
\label{sec:task}

\subsection{Task Definition}

DCASE~2026 Task~5 (ADQA) is a four-way multiple-choice question (MCQ) task over audio clips. Each item pairs one audio recording, one question, and four options with a single correct choice. Every question must be \emph{audio-dependent}: the correct answer cannot be recovered from the text alone.

\subsection{Audio-Dependency Filtering (ADF)}

\begin{figure}[t]
  \centering
  \includegraphics[width=\columnwidth]{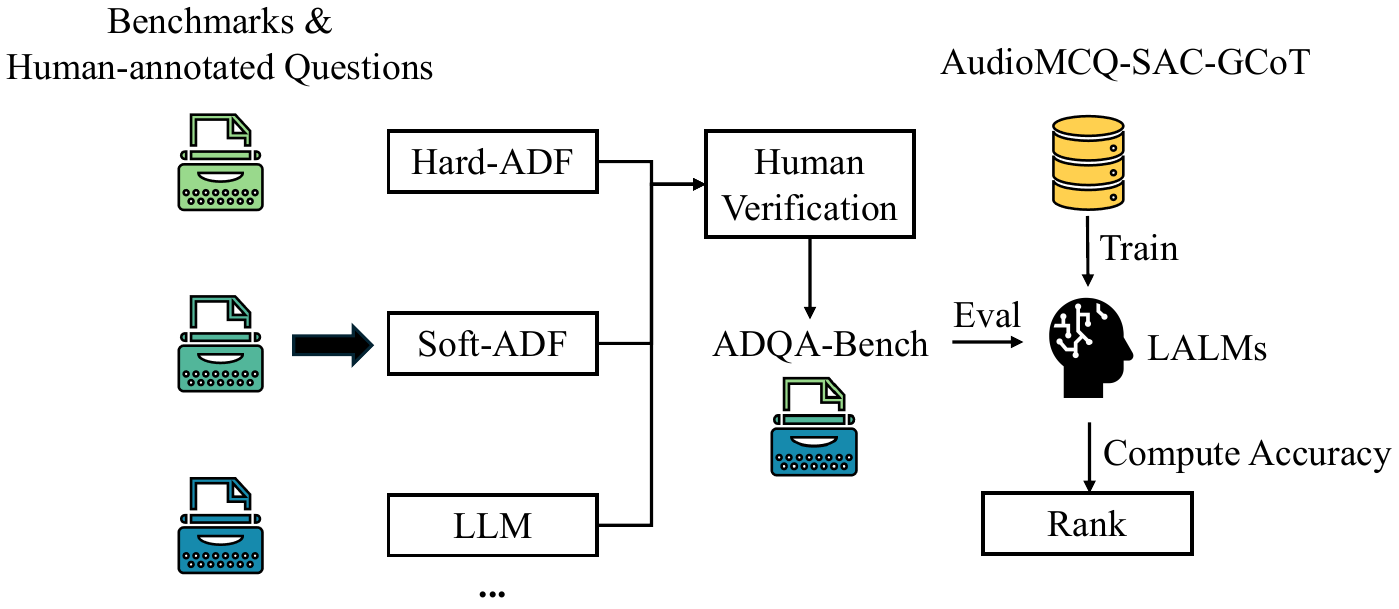}
  \vspace{-18pt}
  \caption{The ADF pipeline. Three filtering stages (Hard-ADF,
  Soft-ADF, large language model (LLM) commonsense) and a final human verification step
  yield ADQA-Bench from existing benchmarks and newly annotated
  multiple-choice-question (MCQ) items; training items follow the
  related Strong Audio-Contribution (StrongAC) selection on the
  AudioMCQ corpus~\cite{he2025audiomcq}, with chain-of-thought (CoT)
  rationales; the figure's AudioMCQ-SAC-GCoT block denotes this
  AudioMCQ-StrongAC-GeminiCoT training corpus (SAC = StrongAC;
  GCoT = Gemini-distilled CoT).}
  \label{fig:adf}
\end{figure}

The ADF cascade~\cite{he2025audiomcq} is a four-stage pipeline (\cref{fig:adf}) that keeps only items whose correct answer depends on the audio signal.

\textbf{Hard-ADF (silent-audio probe).} We replace the audio with silence and ask a pool of LALMs to answer. An item is discarded if more than one blind model still picks the correct option, since the answer is then largely recoverable without listening.

\textbf{Soft-ADF (perplexity filter).} We score each option with language-model perplexity under silent audio. If the correct option has the lowest perplexity, the item is dropped: a text-only preference for the right answer exposes a linguistic shortcut.

\textbf{LLM commonsense filter.} A text-only large language model (LLM) sees the question and options without the audio. Items it answers correctly from commonsense or world knowledge are pruned.

\textbf{Human verification.} A two-pass human review checks that the labeled answer is correct given the audio and that every distractor is plausible but acoustically wrong.

\subsection{Datasets}

\textbf{Training set.} The official training corpus, AudioMCQ-StrongAC-GeminiCoT (hereafter AudioMCQ-StrongAC), is derived from the ${\sim}$571k-item AudioMCQ corpus~\cite{he2025audiomcq}. We keep a subset of the Strong Audio-Contribution (StrongAC) split and add chain-of-thought (CoT) rationales distilled from Gemini~3.1~Pro.

\textbf{Development set.} 1607 MCQ items drawn from existing benchmarks (MMAU~\cite{sakshi2025mmau}, MMAR~\cite{ma2025mmar}, MMSU~\cite{wang2025mmsu}) together with newly annotated questions, all passing the four-stage ADF.

\textbf{Evaluation set.} The hidden ADQA-Bench holds 3000 MCQ items filtered under the same protocol. Team-submission scores in this paper refer to this evaluation set, whereas organizer baseline numbers refer to the development set alone (\cref{sec:dev_eval}).

\textbf{Protocol.} Two tracks are defined by total parameter count: Overall (${\le}\bn{100}$ total, no component $>\bn{30}$) and Lightweight ($<\bn{10}$). Submissions must use open-source LALMs. Manual annotation of evaluation items is forbidden: teams cannot label evaluation audio by hand, re-transcribe it for training, or feed per-item human answers back into their systems. Each team can submit up to four systems.

\subsection{Baseline Systems}

The organizers release five LALM baselines with tuned prompts and score them by top-1 accuracy on the development set. \cref{tab:baselines} reports these scores next to a random-guess level of \pct{25.46}. This exceeds \pct{25} likely because option-length and distractor artifacts make a few choices slightly more likely than a uniform guess. Qwen3-Omni-30B scores highest among the baselines.

\begin{table}[t]
\centering
\caption{Top-1 accuracy of organizer baselines on the development
set. Random-guess level: \pct{25.46}.}
\label{tab:baselines}
\begin{tabular}{lr}
    \toprule
    \textbf{Baseline system} & \textbf{Accuracy} \\
    \midrule
    Fun-Audio-Chat-8B~\cite{funaudio_2025}    & \pct{56.81} \\
    Kimi-Audio-7B~\cite{kimiaudio_2025}        & \pct{46.36} \\
    MiMo-Audio-7B~\cite{mimoaudio_2025}        & \pct{54.57} \\
    Qwen3-Omni-30B~\cite{qwen3omni_2025}       & \textbf{\pct{62.48}} \\
    Step-Audio-2 Mini~\cite{stepaudio2_2025}    & \pct{50.53} \\
    \midrule
    Random guess         & \pct{25.46} \\
    \bottomrule
\end{tabular}
\end{table}

\section{Submission Overview}
\label{sec:overview}

The first edition of DCASE~2026 Task~5 draws 14 teams and 36 ranked
submissions, spanning \pct{31.87} to \pct{58.33} accuracy. Because
each system is scored on 3000 items, gaps of about a percentage
point should be read as ties. Every system
is built on one of six public LALMs, so teams differ mainly in their
backbone choice and how they curate data around it.

The overall champion is \textbf{\sys{Lim\_CAU\_4}} from Chung-Ang
University~\cite{Lim_CAU_t5_2026}, which reaches \pct{58.33} by
ensembling MOSS-Audio-8B-Thinking~\cite{moss_audio_2026} with
Qwen3-Omni-30B~\cite{qwen3omni_2025} (\bn{96.0} in the challenge
metadata; the team reports a \bn{92} effective total over four
MOSS-8B and two Qwen3-Omni-30B inference views).
Its pipeline chains audio-vs-silent difficulty binning,
Group reward-Decoupled Normalization Policy Optimization (GDPO)~\cite{liu2026gdpo}-based Low-Rank Adaptation
(LoRA)~\cite{hu2022lora}, an inference-time acoustic tagger, and
choice-permutation ensembling (see \cref{sec:champion}).
\team{Nam\_IND}~\cite{Nam_IND_t5_2026} applies 4-bit quantized LoRA
(QLoRA)~\cite{dettmers2023qlora} supervised fine-tuning (SFT) with a response normalizer on
Qwen3-Omni-30B (\bn{60}, \pct{57.17}).
\team{Hu\_IOA}~\cite{Hu_IOA_t5_2026} runs MOSS-Audio-8B-Thinking with
zero-shot label-only decoding and 5-view choice-permutation majority
voting (\pct{57.03}). \team{Yin\_XJTLU}~\cite{Yin_XJTLU_t5_2026} builds
a training-free Qwen3-Omni-30B prompt-program ensemble (\pct{56.00}).
\team{Cheng\_Surrey}~\cite{Cheng_Surrey_t5_2026} adds retrieval
augmentation to MOSS-Audio-8B-Thinking (\pct{53.93}), and
\team{Zhang\_WHU}~\cite{Zhang_WHU_t5_2026} combines SFT with
reinforcement learning (RL) on the same backbone (\pct{51.57}). The
remaining eight
teams~\cite{Tathe_UIUC_t5_2026,Huang_JAIST_t5_2026,Kim_SGU_t5_2026,Wu_XMU_t5_2026,ZC_Inst_t5_2026,Guan_HEU_t5_2026,Song_BIT_t5_2026,Xu_HUST_t5_2026}
span from \pct{31.87} to \pct{50.60}.

% All 36 ranked submissions, sorted by Eval accuracy.
% \dag marks systems whose total size is <10B (lightweight track).
\begin{table*}[t]
\centering
\setlength{\tabcolsep}{2pt}
\renewcommand{\arraystretch}{0.9}
% Table-local: make the mono team/system font inherit the surrounding
% table size instead of the body's fixed 8.5pt, so the Submission column
% matches the other columns. Scoped to this table; body text is unchanged.
\renewcommand{\team}[1]{{\fontfamily{lmtt}\selectfont #1}}
\renewcommand{\sys}[1]{{\fontfamily{lmtt}\selectfont #1}}
\caption{All 36 ranked submissions on the DCASE 2026 Task~5 evaluation set, sorted by Eval accuracy. The overall champion is shown in bold; \textdagger\ marks lightweight systems whose total system size is below \bn{10} parameters. Dev/Eval = development/evaluation-set top-1 accuracy; Diff = Eval $-$ Dev in percentage points (pp); -- marks submissions with no published development score. Paradigm is a team-level label from each team's best system, shared by all its submissions. Paradigm abbreviations: H = Hybrid, S = SFT only, S+R = SFT+RL, RL = RL only, TF = Training-free, RAG = Retrieval-augmented generation. Backbone abbreviations: MOSS-8B-Think = MOSS-Audio-8B-Thinking; Qwen3-Omni-30B = Qwen3-Omni-30B-A3B-Instruct; Qwen3-Emb = Qwen3-Embedding-0.6B; MiMo-Audio-7B = MiMo-Audio-7B-Instruct; MOSS-4B-Instruct = MOSS-Audio-4B-Instruct. The \bn{80.0} figure for \sys{Lim\_CAU\_3} follows the challenge-metadata replicated-view convention for a single \bn{8} MOSS-Audio-8B-Thinking checkpoint used across ten permutation views (\bn{8} unique weights); the \bn{96.0} for \sys{Lim\_CAU\_4} likewise follows the metadata, while its report states a \bn{92} effective total over six inference views. \textsuperscript{o}: oracle upper bound for the category-aware audio-attention setting reported by \team{Yin\_XJTLU}; the deployed system's calibrated dev accuracy is not separately reported.}
\label{tab:teams}
\scriptsize
\begin{tabular}{rllllrrrr}
\toprule
Rank & Submission & Institution & Paradigm & Base LALM & Params & Dev (\%) & Eval (\%) & Diff (\%) \\
\midrule
\textbf{1}  & \textbf{\sys{Lim\_CAU\_4}}~\cite{Lim_CAU_t5_2026}                       & \textbf{Chung-Ang U.}      & \textbf{H}  & \textbf{MOSS-8B-Think + Qwen3-Omni-30B} & \textbf{\bn{96.0}} & \textbf{70.50} & \textbf{58.33} & \textbf{$-$12.17} \\
2  & \sys{Lim\_CAU\_3}~\cite{Lim_CAU_t5_2026}                                         & Chung-Ang U.          & H   & MOSS-8B-Think                      & \bn{80.0} & 70.01 & 58.10 & $-$11.91 \\
3  & \sys{Lim\_CAU\_1}~\cite{Lim_CAU_t5_2026}\textsuperscript{\textdagger}            & Chung-Ang U.          & H   & MOSS-8B-Think                      & \bn{8.0}  & 69.63 & 57.30 & $-$12.33 \\
4  & \sys{Nam\_IND\_2}~\cite{Nam_IND_t5_2026}                                         & Independent           & S   & Qwen3-Omni-30B                     & \bn{60.0} & --    & 57.17 & --     \\
5  & \sys{Nam\_IND\_4}~\cite{Nam_IND_t5_2026}                                         & Independent           & S   & Qwen3-Omni-30B + Gemma-4-E4B-it    & \bn{94.0} & --    & 57.13 & --     \\
6  & \sys{Lim\_CAU\_2}~\cite{Lim_CAU_t5_2026}\textsuperscript{\textdagger}            & Chung-Ang U.          & H   & MOSS-8B-Think                      & \bn{8.0}  & 69.70 & 57.07 & $-$12.63 \\
7  & \sys{Hu\_IOA\_4}~\cite{Hu_IOA_t5_2026}\textsuperscript{\textdagger}             & IOA, CAS              & H   & MOSS-8B-Think + Qwen3-Emb          & \bn{8.6}  & 67.70 & 57.03 & $-$10.67 \\
8  & \sys{Nam\_IND\_3}~\cite{Nam_IND_t5_2026}                                         & Independent           & S   & Qwen3-Omni-30B                     & \bn{60.0} & --    & 56.73 & --     \\
9  & \sys{Hu\_IOA\_3}~\cite{Hu_IOA_t5_2026}\textsuperscript{\textdagger}             & IOA, CAS              & H   & MOSS-8B-Think + Qwen3-Emb          & \bn{8.6}  & 66.02 & 56.70 & $-$9.32  \\
10 & \sys{Yin\_XJTLU\_1}~\cite{Yin_XJTLU_t5_2026}                                       & XJTLU                 & TF  & Qwen3-Omni-30B                     & \bn{30.0} & 68.33 & 56.00 & $-$12.33 \\
11 & \sys{Nam\_IND\_1}~\cite{Nam_IND_t5_2026}                                         & Independent           & S   & Qwen3-Omni-30B                     & \bn{60.0} & 67.27 & 55.90 & $-$11.37 \\
12 & \sys{Yin\_XJTLU\_4}~\cite{Yin_XJTLU_t5_2026}\textsuperscript{\textdagger}          & XJTLU                 & TF  & MOSS-8B-Think                      & \bn{8.0}  & 66.40 & 55.80 & $-$10.60 \\
13 & \sys{Yin\_XJTLU\_2}~\cite{Yin_XJTLU_t5_2026}\textsuperscript{\textdagger}          & XJTLU                 & TF  & MOSS-8B-Think                      & \bn{8.0}  & 64.97 & 55.60 & $-$9.37  \\
14 & \sys{Yin\_XJTLU\_3}~\cite{Yin_XJTLU_t5_2026}\textsuperscript{\textdagger}          & XJTLU                 & TF  & MOSS-8B-Think                      & \bn{8.0}  & 66.46\textsuperscript{o} & 55.27 & $-$11.19 \\
15 & \sys{Cheng\_Surrey\_1}~\cite{Cheng_Surrey_t5_2026}\textsuperscript{\textdagger}       & Surrey                & RAG & MOSS-8B-Think                      & \bn{8.0}  & 65.07 & 53.93 & $-$11.14 \\
16 & \sys{Cheng\_Surrey\_2}~\cite{Cheng_Surrey_t5_2026}\textsuperscript{\textdagger}       & Surrey                & RAG & MOSS-8B-Think                      & \bn{8.0}  & 65.01 & 53.50 & $-$11.51 \\
17 & \sys{Zhang\_WHU\_1}~\cite{Zhang_WHU_t5_2026}\textsuperscript{\textdagger}          & WHU + CUHK-SZ         & S+R & MOSS-8B-Think                      & \bn{8.0}  & 62.79 & 51.57 & $-$11.22 \\
18 & \sys{Zhang\_WHU\_2}~\cite{Zhang_WHU_t5_2026}\textsuperscript{\textdagger}          & WHU + CUHK-SZ         & S+R & MOSS-8B-Think                      & \bn{8.0}  & 62.79 & 51.13 & $-$11.66 \\
19 & \sys{Tathe\_UIUC\_1}~\cite{Tathe_UIUC_t5_2026}\textsuperscript{\textdagger}         & UIUC + CMU            & S+R & Qwen2.5-Omni-7B                    & \bn{7.0}  & 58.43 & 50.60 & $-$7.83  \\
20 & \sys{Hu\_IOA\_2}~\cite{Hu_IOA_t5_2026}\textsuperscript{\textdagger}             & IOA, CAS              & H   & Qwen2.5-Omni-7B + Qwen3-Emb        & \bn{7.6}  & 58.93 & 49.87 & $-$9.06  \\
21 & \sys{Tathe\_UIUC\_2}~\cite{Tathe_UIUC_t5_2026}\textsuperscript{\textdagger}         & UIUC + CMU            & S+R & Qwen2.5-Omni-7B                    & \bn{7.0}  & 57.31 & 49.83 & $-$7.48  \\
22 & \sys{Hu\_IOA\_1}~\cite{Hu_IOA_t5_2026}\textsuperscript{\textdagger}             & IOA, CAS              & H   & Qwen2.5-Omni-7B + Qwen3-Emb        & \bn{7.6}  & 58.93 & 49.63 & $-$9.30  \\
23 & \sys{Huang\_JAIST\_1}~\cite{Huang_JAIST_t5_2026}\textsuperscript{\textdagger}        & JAIST                 & S   & Fun-Audio-Chat-8B                  & \bn{8.0}  & 64.84 & 49.60 & $-$15.24 \\
24 & \sys{Kim\_SGU\_4}~\cite{Kim_SGU_t5_2026}\textsuperscript{\textdagger}            & Sogang U.             & H   & MiMo-Audio-7B                      & \bn{7.0}  & --    & 49.13 & --     \\
25 & \sys{Kim\_SGU\_3}~\cite{Kim_SGU_t5_2026}\textsuperscript{\textdagger}            & Sogang U.             & H   & MiMo-Audio-7B                      & \bn{7.0}  & 59.43 & 48.87 & $-$10.56 \\
26 & \sys{Tathe\_UIUC\_3}~\cite{Tathe_UIUC_t5_2026}\textsuperscript{\textdagger}         & UIUC + CMU            & S+R & Qwen2.5-Omni-7B                    & \bn{7.0}  & 56.32 & 48.67 & $-$7.65  \\
27 & \sys{Tathe\_UIUC\_4}~\cite{Tathe_UIUC_t5_2026}\textsuperscript{\textdagger}         & UIUC + CMU            & S+R & Qwen2.5-Omni-7B                    & \bn{7.0}  & 55.51 & 48.33 & $-$7.18  \\
28 & \sys{Kim\_SGU\_2}~\cite{Kim_SGU_t5_2026}\textsuperscript{\textdagger}            & Sogang U.             & H   & MiMo-Audio-7B                      & \bn{7.0}  & 73.43 & 47.97 & $-$25.46 \\
29 & \sys{Wu\_XMU\_1}~\cite{Wu_XMU_t5_2026}\textsuperscript{\textdagger}             & Xiamen + THU          & TF  & Qwen2.5-Omni-7B                    & \bn{7.0}  & 53.90 & 47.20 & $-$6.70  \\
30 & \sys{Kim\_SGU\_1}~\cite{Kim_SGU_t5_2026}\textsuperscript{\textdagger}            & Sogang U.             & H   & MiMo-Audio-7B                      & \bn{7.0}  & 72.76 & 46.77 & $-$25.99 \\
31 & \sys{ZC\_Inst\_1}~\cite{ZC_Inst_t5_2026}\textsuperscript{\textdagger}            & HDU                   & S   & Fun-Audio-Chat-8B                  & \bn{8.0}  & 64.45 & 46.03 & $-$18.42 \\
32 & \sys{Guan\_HEU\_1}~\cite{Guan_HEU_t5_2026}\textsuperscript{\textdagger}           & HEU + UTS             & S   & Fun-Audio-Chat-8B                  & \bn{8.0}  & 53.08 & 43.37 & $-$9.71  \\
33 & \sys{Song\_BIT\_1}~\cite{Song_BIT_t5_2026}\textsuperscript{\textdagger}           & BIT                   & RL  & MiMo-Audio-7B                      & \bn{7.0}  & 50.53 & 43.33 & $-$7.20  \\
34 & \sys{Song\_BIT\_2}~\cite{Song_BIT_t5_2026}\textsuperscript{\textdagger}           & BIT                   & RL  & MiMo-Audio-7B                      & \bn{7.0}  & 48.10 & 41.53 & $-$6.57  \\
35 & \sys{Song\_BIT\_3}~\cite{Song_BIT_t5_2026}\textsuperscript{\textdagger}           & BIT                   & RL  & MiMo-Audio-7B                      & \bn{7.0}  & 54.20 & 40.97 & $-$13.23 \\
36 & \sys{Xu\_HUST\_1}~\cite{Xu_HUST_t5_2026}\textsuperscript{\textdagger}            & HUST                  & S   & MOSS-4B-Instruct + DeBERTa-v3      & \bn{5.4}  & 56.69 & 31.87 & $-$24.82 \\
\bottomrule
\end{tabular}
\end{table*}

\begin{figure}[t]
  \centering
  \includegraphics[width=0.74\columnwidth]{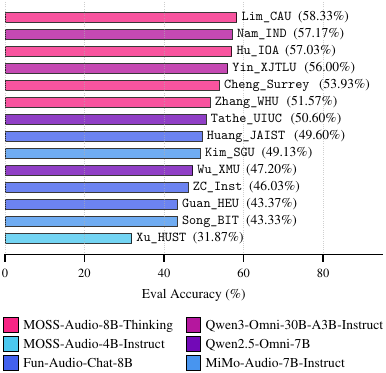}
  \caption{Best eval-set accuracy per team, color-coded by primary
    base LALM. The dashed line marks the \pct{49.65} mean across the
    14 teams' best submissions; four teams reach at least \pct{56}.}
  \label{fig:ranking}
\end{figure}

\Cref{tab:teams} lists all 36 ranked submissions, with \textdagger\
marking lightweight-eligible systems ($<\bn{10}$); \cref{fig:ranking}
plots each team's best accuracy against the lightweight bar. The
lightweight champion is \textbf{\sys{Lim\_CAU\_1}} at \pct{57.30},
\pp{1.03} behind the \bn{96} ensemble.

MOSS-Audio-8B-Thinking is the most common backbone at the top of the leaderboard. Four of
14 teams (\team{Lim\_CAU}, \team{Hu\_IOA}, \team{Cheng\_Surrey},
\team{Zhang\_WHU}) pick it as the backbone of their best
submission, and \team{Yin\_XJTLU} uses it in three of four submissions. Inside \team{Hu\_IOA}, an
un-fine-tuned MOSS run reaches \pct{57.03} while their tuned
Qwen2.5-Omni-7B~\cite{qwen25omni_2025,Hu_IOA_t5_2026} scores
\pct{49.87}; this pair confounds base model, size, and tuning status,
so we treat it as one data point. Without controlled ablations we cannot separate the
contributions of checkpoint alignment, LoRA~\cite{hu2022lora}
compatibility (\team{Lim\_CAU}, \team{Zhang\_WHU}), and
retrieval support (\team{Cheng\_Surrey}).

Among the top five, \team{Yin\_XJTLU} favors MOSS overall, yet its best submission uses a different backbone. Its
Qwen3-Omni prompt-program ensemble (\sys{Yin\_XJTLU\_1}, \pct{56.00})
narrowly beats its own MOSS-Audio-8B-Thinking variants: audio-token attention biases (\sys{Yin\_XJTLU\_2}/\sys{Yin\_XJTLU\_3}) and an option-elimination prompt program (\sys{Yin\_XJTLU\_4}). Later
sections analyze these submissions by paradigm,
backbone, and common techniques (\cref{sec:methods}), and by their performance patterns (\cref{sec:analysis}).

\section{Methodologies and Trends}
\label{sec:methods}

We compare the 36 submissions by backbone and technique, and the 14
teams by training paradigm---one label per team, taken from its best system and applied to all of that team's submissions. Paradigm
comparisons are descriptive only: cell sizes (1--5 teams) are too
small for statistical inference, and paradigm is closely confounded with each team's backbone choice.

\subsection{Paradigm Taxonomy}

Each team's best system falls into one of six paradigms
(see \cref{tab:teams}), reported as best, mean, and team count (best omitted where not informative):
hybrid~(\pct{58.33}, \pct{54.83}, 3), training-free~(\pct{56.00},
\pct{51.60}, 2), supervised fine-tuning with reinforcement learning
(SFT+RL; mean \pct{51.09}, 2 teams), SFT only (mean \pct{45.61},
5 teams), and RL only and retrieval-augmented (1 team each).
\textbf{Hybrid} systems fuse components from different
paradigms~\cite{Lim_CAU_t5_2026,Hu_IOA_t5_2026,Kim_SGU_t5_2026}
and achieve the highest best accuracy and mean. Strong SFT-only
(\team{Nam\_IND}) and training-free (\team{Yin\_XJTLU}) systems
still land in the \pct{55}--\pct{57} band, within \pp{2.4} of the
top hybrid system.

\subsection{Base Model Choice}

\cref{tab:basemodel} groups submissions by primary backbone. Two
models dominate. MOSS-Audio-8B-Thinking~\cite{moss_audio_2026}
(13 submissions, 5 teams) gives the best single-submission accuracy
(\pct{58.33}, from a system that also fuses Qwen3-Omni-30B views) and a backbone mean of \pct{55.49}.
Qwen3-Omni-30B~\cite{qwen3omni_2025} (5 submissions, 2 teams) has
a slightly higher raw mean (\pct{56.59}), but four of its five
submissions come from one team's SFT variants, so the cross-team
mean is not meaningful at 2 teams. All other backbones plateau below
\pct{51}~\cite{Tathe_UIUC_t5_2026,Wu_XMU_t5_2026,Huang_JAIST_t5_2026,Guan_HEU_t5_2026,ZC_Inst_t5_2026,Song_BIT_t5_2026,Xu_HUST_t5_2026}.
In the one available within-team comparison, added
system size gives only marginal gains once a strong \bn{8} backbone
is in place: \team{Lim\_CAU}'s \bn{96} ensemble beats its
lightweight entry by only \pp{1.03}~\cite{Lim_CAU_t5_2026}. A single comparison cannot isolate the effect of size.

% Per-base-model submission statistics on the evaluation set.
\begin{table}[t]
\centering
\setlength{\tabcolsep}{4pt}
\renewcommand{\arraystretch}{1.05}
\caption{Submissions grouped by primary base LALM (first model listed in each submission's meta file). Sorted by the best evaluation-set accuracy attained with each backbone. Abbreviations: MOSS-8B-Think = MOSS-Audio-8B-Thinking; Qwen3-Omni-30B = Qwen3-Omni-30B-A3B-Instruct; MOSS-4B-Instruct = MOSS-Audio-4B-Instruct; MiMo-Audio-7B~=~\mbox{MiMo-Audio-7B-Instruct}.}
\label{tab:basemodel}
\small
\begin{tabular}{lrrrr}
\toprule
Base Model & \#Sub. & \#Teams & Best (\%) & Mean (\%) \\
\midrule
MOSS-8B-Think         & 13 & 5 & 58.33 & 55.49 \\
Qwen3-Omni-30B        &  5 & 2 & 57.17 & 56.59 \\
Qwen2.5-Omni-7B       &  7 & 3 & 50.60 & 49.16 \\
Fun-Audio-Chat-8B     &  3 & 3 & 49.60 & 46.33 \\
MiMo-Audio-7B         &  7 & 2 & 49.13 & 45.51 \\
MOSS-4B-Instruct      &  1 & 1 & 31.87 & 31.87 \\
\bottomrule
\end{tabular}
\end{table}

\subsection{Commonly Used Techniques}

Across the 36 submissions, most teams share a small pool of
techniques on a common backbone. \cref{tab:methodology}
lists the 16 technique
tags used by at least two teams. Prompt engineering (11 teams),
LoRA~\cite{hu2022lora} (9), SFT (9), and CoT prompting (9) are
the most common; Group Relative Policy Optimization (GRPO)~\cite{shao2024grpo} appears in 5 teams and
majority voting in seven teams.

% Methodology tags aggregated from team technical-report summaries.
% Two-column layout to save vertical space.
\begin{table}[t]
\centering
\setlength{\tabcolsep}{2pt}
\renewcommand{\arraystretch}{0.95}
\caption{Methodology tags adopted by at least two teams (16 tags total). ``Top team'' is the best-ranked team that uses the tag in any of its submissions, not necessarily in that team's best-scoring one, and is not a performance attribution for the tag. Abbreviations: LoRA = Low-Rank Adaptation; SFT = supervised fine-tuning; GRPO = Group Relative Policy Optimization; GDPO = Group reward-Decoupled \mbox{Normalization Policy Optimization}.}
\label{tab:methodology}
\footnotesize
\begin{tabular}{lcl@{\hspace{6pt}}lcl}
\toprule
Tag & \#T & Top team & Tag & \#T & Top team \\
\midrule
Prompt Engineering  & 11 & \team{Lim\_CAU}     & Self-Consistency    & 3 & \team{Yin\_XJTLU} \\
Chain-of-Thought    & 9  & \team{Hu\_IOA}      & Voting Ensemble     & 3 & \team{Lim\_CAU}   \\
LoRA                & 9  & \team{Lim\_CAU}     & Choice-Permut. Ens. & 3 & \team{Lim\_CAU}   \\
SFT                 & 9  & \team{Nam\_IND}     & Reward Model        & 3 & \team{Zhang\_WHU} \\
Majority Voting     & 7  & \team{Lim\_CAU}     & Knowledge Distill.  & 2 & \team{Zhang\_WHU} \\
GRPO                & 5  & \team{Hu\_IOA}      & GDPO                & 2 & \team{Lim\_CAU}   \\
Hard-Example Mining & 4  & \team{Lim\_CAU}     & LoRA (4-bit)        & 2 & \team{Nam\_IND}   \\
Test-Time Scaling   & 4  & \team{Lim\_CAU}     & Prompt Calibration  & 2 & \team{Lim\_CAU}   \\
\bottomrule
\end{tabular}
\end{table}

\textit{Choice-permutation ensembling}:
\team{Lim\_CAU}~\cite{Lim_CAU_t5_2026} fuses MOSS and Qwen3-Omni
views via weighted log-probs ($\lambda{\approx}0.46$);
\team{Yin\_XJTLU}~\cite{Yin_XJTLU_t5_2026} votes across five prompt
programs; \team{Zhang\_WHU}~\cite{Zhang_WHU_t5_2026} gains
${\sim}$\pp{3.8} from eight-sample voting atop
GRPO~\cite{shao2024grpo}.
\textit{RL objectives}: five teams train with
GRPO~\cite{Hu_IOA_t5_2026,Zhang_WHU_t5_2026,Tathe_UIUC_t5_2026,Kim_SGU_t5_2026,Song_BIT_t5_2026},
and \team{Lim\_CAU} applies GDPO-style reward normalization on LoRA
adapters~\cite{Lim_CAU_t5_2026}.
\textit{Reward engineering}: atop RL, \team{Hu\_IOA} combines
accuracy, format, question-type, embedding-similarity, and
length-regularization rewards, optimized with GDPO~\cite{Hu_IOA_t5_2026}, while
\team{Song\_BIT} adds an attention-anchoring reward that ties
deep-layer cross-attention to audio key positions~\cite{Song_BIT_t5_2026}.
\textit{Acoustic-feature injection}: \team{Lim\_CAU}'s
router~\cite{Lim_CAU_t5_2026} matches question keywords to
categories---speech, speaker, music/scene---and runs only the
relevant taggers: voice activity detection (VAD),
Whisper~\cite{radford2023whisper} prosody, pyannote
diarization~\cite{bredin2023pyannote},
ECAPA-TDNN~\cite{desplanques2020ecapa} speaker verification, and
Contrastive Language-Audio Pretraining (CLAP)~\cite{wu2023clap}
scene matching, injecting the outputs as natural-language snippets.
\textit{SFT on AudioMCQ-StrongAC}: the dominant training
recipe~\cite{Guan_HEU_t5_2026,Huang_JAIST_t5_2026,Tathe_UIUC_t5_2026,Hu_IOA_t5_2026,Zhang_WHU_t5_2026},
with targets from answer text to structured or \mbox{teacher-rewritten} CoT rationales.

\subsection{Winning System: \team{Lim\_CAU}}
\label{sec:champion}

The winning recipe~\cite{Lim_CAU_t5_2026} scores each training
candidate under matched (audio, silent) prompts and bins them into
six difficulty classes. From 167{,}541
AudioMCQ-StrongAC~\cite{he2025audiomcq} candidates, 12{,}800 rows
are drawn with group-level quotas. A choice-layout calibration
oversamples option~D targets (A/B/C/D$\,{=}\,$1.0/1.0/1.0/1.3) to
counter reported positional bias in MOSS outputs.
GDPO~\cite{liu2026gdpo}-based LoRA~\cite{hu2022lora} then tunes the
model. At inference, an acoustic tagger injects librosa~\cite{mcfee2015librosa}-derived evidence from the taggers above. The final submission \sys{Lim\_CAU\_4} fuses
MOSS and Qwen3-Omni-30B views via weighted log-probs
($\lambda{\approx}0.46$) and reaches \pct{58.33}, the highest
accuracy of any single submission. No ablation isolates the acoustic tagger from the two-backbone ensemble, so its standalone gain is unknown.

\section{Performance Analysis}
\label{sec:analysis}

\subsection{Question-level Difficulty Distribution}
\label{sec:difficulty}

\begin{figure}[t]
  \centering
  \includegraphics[width=0.58\columnwidth]{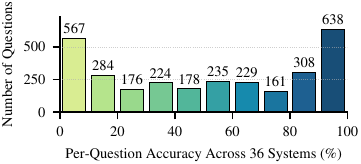}
  \caption{Per-question accuracy across the 36 submissions, grouped
  into ten 10-percentage-point bins. Mass concentrates in the
  left-most (hardest) and right-most (easiest) bins.}
  \label{fig:difficulty}
\end{figure}

\Cref{fig:difficulty} bins the 3000 evaluation questions by the
fraction of submissions that answer correctly. The distribution is
bimodal: 567 questions (\pct{18.9}) are solved by fewer than
\pct{10} of systems and 638 (\pct{21.3}) by more than \pct{90}. 233
items are missed by every ranked system; 110 are solved by all.
For \pct{47.6} of items, fewer than half of the 36 submissions answer correctly, so shared failures reach well past the 233-item
universal-failure set. This points to weaknesses shared by the LALM backbones used here; \cref{sec:discussion} revisits their likely causes in detail.

\subsection{Dev-to-Eval Accuracy Drop}
\label{sec:dev_eval}

The evaluation set is designed to be harder than the development
split~\cite{he2025audiomcq}. Of the 30 submissions with a comparable
development score (excluding \sys{Nam\_IND\_2/3/4} and \sys{Kim\_SGU\_4},
which publish none, and \sys{Song\_BIT\_3} and \sys{Yin\_XJTLU\_3}, whose
reported numbers are an MMAU test-mini proxy and an oracle upper
bound), every one is lower on
eval: mean gap \pp{$-$11.91}, median \pp{$-$10.91}, standard deviation (SD) \pp{5.24},
ranging from \pp{$-$6.6} to \pp{$-$25.99}.
Several teams contribute multiple submissions sharing a backbone
and training data, so $n{=}30$ is small and we treat the SD descriptively.

The two teams with the largest gaps are \team{Kim\_SGU} at \pp{$-$25.99}
\cite{Kim_SGU_t5_2026} and \team{Xu\_HUST} at \pp{$-$24.82}
\cite{Xu_HUST_t5_2026}; both of which tune adapter and leaf selection using development-set analysis, consistent with development-set overfitting. The smallest gaps come from training-free
or RL-heavy systems whose fine-tuning data excludes
development-benchmark items: \team{Song\_BIT} at \pp{$-$6.6}/\pp{$-$7.2}
\cite{Song_BIT_t5_2026}, \team{Wu\_XMU} at \pp{$-$6.7}
\cite{Wu_XMU_t5_2026}, and \team{Tathe\_UIUC} at \pp{$-$7.2} to
\pp{$-$7.83} \cite{Tathe_UIUC_t5_2026}, all with lower development scores.

\subsection{Reasoning-Time Interventions}
\label{sec:cot_hurts}

CoT-target SFT raises development accuracy for some systems---\team{Hu\_IOA}'s
Qwen2.5-Omni from \pct{55.88} to \pct{57.93}~\cite{Hu_IOA_t5_2026} and
\team{Zhang\_WHU}'s MOSS-Audio from \pct{52.21} to
\pct{56.94}~\cite{Zhang_WHU_t5_2026}---but three other reasoning-time add-ons do not.
A separate training-time case shows the reverse risk: \team{Hu\_IOA} reports a \pp{$-$8.71} development-split drop from SFT on
MOSS-Audio-8B-Thinking, with development accuracy falling from
\pct{67.70} to \pct{58.99}; they attribute the drop to catastrophic
forgetting of the model's native audio reasoning rather than to CoT
supervision itself~\cite{Hu_IOA_t5_2026}.
\team{Nam\_IND} loses \pp{$-$1.93} with a CoT-then-letter
prompt~\cite{Nam_IND_t5_2026}.
\team{Tathe\_UIUC}'s Prometheus~\cite{kim2024prometheus} CoT
judge is statistically tied with its no-judge run
(\pct{57.31} vs \pct{58.43}, McNemar $p{=}0.23$)~\cite{Tathe_UIUC_t5_2026}.
\team{Yin\_XJTLU} drops \pp{$-$8.96} by replacing majority voting with a rule that picks the lowest-entropy branch across its training-free prompt programs~\cite{Yin_XJTLU_t5_2026}. This is a decoding-time change only, so it isolates the selection rule from CoT training.

\section{Discussion and Future Directions}
\label{sec:discussion}

ADQA-Bench makes audio dependence measurable:
submissions span \pct{31.87}--\pct{58.33} on the evaluation set, and 233 items (\pct{7.8}) are missed by every
system (\cref{sec:difficulty}). An oracle picking the best answer
across the 36 submissions solves 2767 of 3000 items (\pct{92.2}),
yet no single system comes close. Fine-tuning on AudioMCQ-StrongAC
derivatives~\cite{he2025audiomcq,Tathe_UIUC_t5_2026,Zhang_WHU_t5_2026,Hu_IOA_t5_2026}
is the most common training approach; \cref{sec:cot_hurts} shows
reasoning-time interventions give mixed results---CoT-target SFT
helps on some backbones, whereas CoT prompting, CoT-based judging,
and entropy-based branch selection do not. Better acoustic grounding
(cf. \cref{sec:methods}) is a next direction, though no submission
isolates its contribution~\cite{Lim_CAU_t5_2026,Song_BIT_t5_2026}.

The dev-to-eval variance (\cref{sec:dev_eval}) motivates a held-out
tuning split, separate from the development set. Because the
development set re-filters
MMAU~\cite{sakshi2025mmau}, MMAR~\cite{ma2025mmar} and
MMSU~\cite{wang2025mmsu} items (\cref{sec:task}), such a split
separates benchmark overlap from overfitting. The 233
universally failed items (\cref{sec:difficulty}) expose capability
gaps common to these LALM backbones, pointing to better audio
encoders and audio-grounded pre-training. The narrow \pp{1.03} gap
between the \bn{96} champion and the \bn{8} lightweight winner shows
data curation and inference-time orchestration rival raw scale.
Extending the benchmark to open-ended generation would blunt
elimination shortcuts but~needs~\mbox{an audio-blind judge.}

\clearpage
\bibliographystyle{IEEEtran}
\def\IEEEbibitemsep{1.2ex plus .1ex minus .1ex}
\bibliography{refs}

\end{document}